\documentclass[pre,preprintnumbers,amsmath,amssymb]{revtex4}
\usepackage{graphicx}
\usepackage{bm}

\begin{document}
\title{ Brownian Functionals in Physics and Computer Science}
\author {Satya N. Majumdar }

\affiliation{ Laboratoire de Physique Th\'eorique et Mod\`eles Statistiques,
        Universit\'e Paris-Sud. B\^at. 100. 91405 Orsay Cedex. France}

\begin{abstract}
This is a brief review on Brownian functionals in one dimension and their various applications, a 
contribution to the special issue ``The Legacy of Albert Einstein" of Current Science. After a brief
description of Einstein's original derivation of the diffusion equation, this article
provides a pedagogical introduction to the path integral methods leading to the
derivation of the celebrated Feynman-Kac formula. The usefulness of this technique
in calculating the statistical properties of Brownian functionals is illustrated
with several examples in physics and probability theory, with particular
emphasis on applications in computer science. The statistical properties of ``first-passage Brownian
functionals" and their applications are also discussed.

\end{abstract}

\maketitle

\date{\today}

\section{introduction}

The year 2005 marks the centenary of the publication of three remarkable papers by Einstein, one on 
Brownian motion~\cite{E1}, one on special 
relativity~\cite{E2}, and the other one on the photoelectric effect and light quanta~\cite{E3}.
Each of them made a revolution on its own.
In particular, his paper on Brownian motion (along with the related work by 
Smoluchowsky~\cite{Smolu} and Langevin~\cite{Langevin}) had a more sustained and 
broader impact,
not just in traditional `natural' sciences such as physics, astronomy, chemistry, biology and mathematics
but even in `man-made' subjects such as economics and computer science. The range of applications
of Einstein's Brownian motion and his theory of diffusion is truely remarkable. The ever emerging
new applications in diverse fields have made the Brownian motion a true legacy and a great gift
of Einstein to science.    

There have been numerous articles in the past detailing the history of Brownian motion prior to and after
Einstein. Reviewing this gigantic amount of work is beyond the scope of this article. This year 
two excellent reviews on the Brownian motion with its history and applications have been
published, one by
Frey and Kroy~\cite{FK} and the other by Duplantier~\cite{Dup}. The former discusses the
applications of Brownian motion in soft matter and biological physics and the latter,
after a very nice historical review, discusses the applications of Brownian motion
in a variety of two dimensional growth problems and their connections to the conformal field theory.  
Apart from these two reviews, there have been numerous other recent reviews on the 100
years of Brownian motion~\cite{others}--it is simply not possible to cite
all of them within the limited scope of this article and I apologise for that.
The purpose of the present article is to discuss some complementary aspects
of Brownian motion that are not covered by the recent reviews mentioned above.
 
After a brief introduction to Einstein's original derivation of the Stokes-Einstein
relation and the diffusion equation in Section II, the principal focus of the rest
of the article will be
on the statistical properties of functionals of one dimensional Brownian motion,
with special emphasis on their applications in physics and computer science.
If $x(\tau)$ represents a Brownian motion, a Brownian functional over a fixed time interval
$[0,t]$ is simply defined as $T=\int_0^t U\left(x(\tau)\right)\,d\tau$, where $U(x)$ is 
some prescribed
arbitrary function. For each realization of the Brownian path, the quantity $T$
has a different value and one is interested in the probability
density function (pdf) of $T$. It was Kac who first realized~\cite{Kac} that the 
statistical properties of one dimensional Brownian functionals can be studied by cleverly using    
the path integral method devised by Feynman in his unpublished Ph.D thesis at Princeton.
This observation of Kac thus took Einstein's classical diffusion process into
yet another completely different domain of physics namely the quantum mechanics
and led to the discovery of the celebrated Feynman-Kac formula.  
Since then Brownian functionals have found numerous applications in diverse
fields ranging from  probability theory~\cite{Kac,Yor1} and finance~\cite{finance}
to disordered systems and mesocopic physics~\cite{CDT}. In this article I will
discuss some of them, along with some recent applications of Brownian 
functionals in computer science.
  
After a brief and pedagogical derivation of the path integral methods leading to the
Feynman-Kac formula in Sections III, I will discuss several applications from
physics, computer science and graph theory in Section IV. In Section V,
the statistical properties of ``first-passage Brownian
functionals" will be discusssed. A first-passage functional is defined as $T =\int_0^{t_f} 
U\left(x(\tau)\right)\,d\tau$ where $t_f$ is the
first-passage time of the Brownian process $x(\tau)$, i.e. the first time the process
crosses zero. Such first-passage functionals have many applications, e.g.
in the distribution of lifetimes of comets, in queueing theory and also in transport
properties in disordered systems. Some of these applications will be discussed in 
Section V.

The diverse and ever emerging new applications of Brownian functionals briefly presented here 
will hopefully convince the reader that `Brownian functionalogy' merits the status of
a subfield of statistical physics (and stochastic calculus) itself and is certainly a part of 
the legacy that Einstein left behind.

\section{Einstein's Theory of Brownian Motion and Langevin's Stochastic Equation}

Einstein's 1905 paper on Brownian motion~\cite{E1} achieved two important milestones: (i) to relate
macroscopic kinetic parameters such as the diffusion constant and friction coefficient 
to the correlation functions characterizing fluctuations of microscopic variables--known
as a fluctuation-dissipation relation and (ii) to provide a derivation of the celebrated
diffusion equation starting from the microscopic irregular motion of a particle--thus laying
the foundation of the important field of ``stochastic processes".

\subsection{A fluctuation-dissipation relation}

Very briefly, Einstein's argument leading to the derivation of fluctuation-dissipation relation goes
as follows. Imagine a dilute gas of noninteracting Brownian particles in a solvent under a constant
volume force $K$ (such as gravity) on each particle. For simplicity, we consider a one dimensional system here, 
though the arguments
can be generalized straightforwardly to higher dimensions. There are two steps to the argument.
The first step is to assume that the dilute gas of Brownian particles suspended in a solvent 
behaves as an ideal gas and hence exerts an osmotic pressure on the container giving rise
to a pressure field. The pressure $p(x)$ at 
point $x$ is related to the density $\rho(x)$ via the equation of state for an ideal gas: 
$p(x)= k_B T \rho(x)$, where $k_B$ is the Boltzmann's constant and $T$ is the temperature. The force per 
unit volume 
due to the pressure field
$-{\partial_x p(x)}$ must be balanced at equilibrium by the net external force density
$K \rho(x)$, leading to the force balance condition: $K\rho(x) = -{\partial_x p(x)}=-k_B\,T \partial_x \rho(x)$.
The solution is simply 
\begin{equation}
\rho(x)= \rho(0)\, \exp\left(-\frac{K}{k_B T}\,x\right).
\label{e1}
\end{equation}

The next step of the argument
consists of identifying two currents in the system. The first is the diffusion current $j_{\rm diff}=-D \partial_x 
\rho(x)$ where $D$ is defined as the diffusion coefficient. The second is the drift
current due
to the external force, $j_{\rm drift}$ which can be computed as follows. Under a constant external force, 
each particle achieves at long times a terminal drift velocity, $v= K/\Gamma$ where $\Gamma$ is the
friction coefficient. For spherical particles of radius $a$, $\Gamma$ is given by the Stoke's
formula, $\Gamma= 6\pi \eta a$ where $\eta$ is the viscosity. Thus, $j_{\rm drift}= v\rho(x)= K\rho(x)/\Gamma$.
Now, at equilibrium, the net current in a closed system must be zero, $j=j_{\rm diff}+ j_{\rm drift}=0$
leading to the equation $-D\partial_x \rho(x)+ K\rho(x)/\Gamma=0$. The solution is
\begin{equation}
\rho(x)= \rho(0)\,\exp\left(-\frac{K}{\Gamma D}\,x\right).
\label{e2}
\end{equation}
Comparing Eqs. (\ref{e1}) and (\ref{e2}) Einstein obtained the important relation
\begin{equation}
D= \frac{k_B\, T}{\Gamma},
\label{e3}
\end{equation} 
which is known today as the Stokes-Einstein relation that connects macroscopic 
kinetic coefficients such as $D$ and $\Gamma$ to the thermal fluctuations characterized
by the temperature $T$.  

\subsection{Diffusion as a microscopic process}

In addition to the fluctuation-dissipation relation in Eq.~(\ref{e3}), Einstein's 1905 paper
on Brownian motion also provided an elegant derivation of the diffusion equation that expressed
the diffusion constant $D$ in terms of microscopic fluctuations. Since the particles are independent,
the density $\rho(x,t)$ can also be interpreted as the probability $\rho(x,t)\equiv P(x,t)$ that
a single Brownian particle is at position $x$ at time $t$ and the aim is to derive an evolution equation
for $P(x,t)$ by following the trajectory of a single particle. Here one assumes that the particle is free, i.e. 
not subjected to any external drift. Einstein considered the particle at position $x$ at time $t$ and asssumed that in a 
microscopic time step $\Delta t$, the particle
jumps by a random amount $\Delta x$ which is thus a stochastic variable. He then wrote down an evolution   
equation for $P(x,t)$
\begin{equation}
P(x,t+\Delta t) = \int_{-\infty}^{\infty} P(x-\Delta x, t)\, \phi_{\Delta t}(\Delta x)\, d (\Delta x)
\label{diff1}
\end{equation}
where $\phi_{\Delta t}(\Delta x)$ is the normalized probability density of the `jump' $\Delta x$ in time 
step $\Delta t$.
This evolution equation 
is known today as the Chapman-Kolmogorov equation and it inherently assumes that the stochastic process
$x(t)$ is Markovian. This means that the jump variables $\Delta x$'s are independent from step to step, so
that the position $x(t)$ of the particle at a given time step depends only on its previous time step
and not on the full previous history of evolution. Next Einstein assumed that $P(x-\Delta x,t)$ in the
integrand in Eq.~(\ref{diff1}) can be Taylor expanded assuming `small' $\Delta x$. This gives
\begin{equation}
P(x,t+\Delta t)=P(x,t) - \mu_1 \frac{\partial P}{\partial x} + \frac{\mu_2}{2!} \frac{\partial^2 P}{\partial 
x^2}+\dots
\label{diff2}
\end{equation}
where $\mu_k = \int_{\infty}^{\infty} (\Delta x)^k \phi_{\Delta t}(\Delta x) d (\Delta x)$ is the $k$-th moment of the 
jump variable $\Delta x$. 
Furthermore, the absence
of external drift sets $\mu_1=0$. Dividing both sides of Eq.~(\ref{diff2}) by $\Delta t$, taking
the limit $\Delta t \to 0$ and keeping only the leading nonzero term (assuming the higher order terms 
vanish as $\Delta t \to 0$) one gets the diffusion equation
\begin{equation}
\frac{\partial P}{\partial t} = D\, \frac{\partial^2 P}{\partial x^2}
\label{diff3}
\end{equation}
where the diffusion constant 
\begin{equation}
D =\lim_{\Delta t \to 0}\, \frac{\mu_2}{2\Delta t}= \lim_{\Delta t \to 0}\, \frac{1}{2\Delta t}\,
\int_{-\infty}^{\infty}(\Delta x)^2\, \phi_{\Delta t}(\Delta x)\, d (\Delta x) = \lim_{\Delta t\to 0}\, 
\frac{\langle (\Delta x)^2\rangle}{2\,\Delta t},
\label{diff4}
\end{equation}
where $\langle (\Delta x)^2\rangle$ is the average of the square of the microscopic displacement in 
a microscopic time step $\Delta t$. 
Thus Einstein was able to express the constant $D$ that appears as a coefficient in the macrosopic diffusion 
current $j_{\rm diff}= - D \partial_x P$ in terms of the microscopic fluctuation $\Delta x$ in the position of
a Brownian particle. 
This derivation also brings out the fundamental principle of the diffusion process, i.e. the length scale
must scale as the square root of the time scale.  

The position of the Brownian particle can evolve via many possible `stochastic' trajectories. The
diffusion equation (\ref{diff3}) describing the evolution of the probability density sums
up the effects of all underlying stochastic trajectories. However, it is often useful to
have a mathematical description of each single trajectory. This brings us to the description
of the diffusion process \`a la Langevin\cite{Langevin}. It is clear from Einstein's derivation that
the local slope of an evolving 
trajectory at time $t$ can be written as 
\begin{equation}
\frac{\Delta x}{\Delta t} = \xi_{\Delta t}(t)
\label{slope1}
\end{equation}
where $\xi_{\Delta t}(t)$ is a random `noise' which is independent from one microscopic step to another,
and it has zero mean. Its variance at a given time $t$, in the continuum limit $\Delta t\to 0$, can also
be computed from Eq.~(\ref{diff4}).
One gets $\langle \xi_{\Delta t}^2(t)\rangle = {\langle (\Delta x)^2\rangle}/{(\Delta t)^2}=
2D/\Delta t$ as $\Delta t\to 0$. Thus the noise term typically scales as $1/\sqrt{\Delta t}$
as $\Delta t\to 0$. The correlation function of the noise between two different
times can then be written as, 
\begin{eqnarray}
\langle \xi_{\Delta t}(t)\xi_{\Delta t}(t')\rangle &=& 0\quad\quad\quad {\rm if}\,\,\, t\ne t' \nonumber \\
&=& \frac{2D}{\Delta t} \quad\quad {\rm if}\,\,\, t=t'
\label{corr1}
\end{eqnarray}
In the continuum limit $\Delta t \to 0$, the noise $\xi_{\Delta t}(t)$ then tends to 
a limiting noise $\xi(t)$ which has zero mean and a correlator, 
$\langle \xi(t)\xi(t')\rangle = 2D \delta(t-t')$. This last result follows
by formally taking the limit $\Delta t \to 0$ in Eq.~(\ref{corr1}) where, loosely
speaking, one replaces the $1/\Delta t$ by $\delta(0)$.
Such a noise is called a `white' noise.  
Thus, in the continuum limit $\Delta t\to 0$,  Eq. (\ref{slope1}) reduces
to the celebrated Langevin equation,
\begin{equation}
\frac{dx}{dt} = \xi(t)
\label{lange1}
\end{equation}
where $\xi(t)$ is a white noise.
Moreover, in the continuum limit $\Delta t\to 0$, one can assume, without any loss of generality, 
that the white noise $\xi(t)$ is Gaussian. This means that 
the joint probability distribution of a particular
history of the noise variables $\left[ \{\xi(\tau)\}, {\rm for}\,\, 0\le \tau\le t\right]$ can
be written as
\begin{equation}
{\rm Prob}\left[\{\xi(\tau)\}\right] \propto \exp\left[-\frac{1}{4D}\int_0^{t}  \xi^2(\tau) d\tau\right].
\label{pi0}
\end{equation}
We will see later that this particular fact plays the key role in the representation of Brownian motion as a
path integral. The Brownian motion $x(t)$ can thus be represented as the integrated white noise,
$x(t)=x(0)+ \int_0^{t} \xi(\tau) d\tau$. While the physicists call this a Brownian motion,
the mathematicians call this integrated white noise the Wiener process, named after
the mathematician N. Wiener.

Langevin's formulation in Eq.~(\ref{lange1}) also makes a correspondence between Brownian motion
and the random walk problem where the position $x_n$ of a random walker after $n$ steps evolves via
\begin{equation}
x_n = x_{n-1} + \xi_n
\label{rw1}
\end{equation}
where $\xi_n$'s are independent random variables, each drawn from the common 
distribution $\phi(\xi)$
for each step $n$.
In fact, the idea of understanding Brownian motion in terms of random walks
was first conceived by Smoluchowsky~\cite{Smolu}. The Langevin equation representation of Brownian motion 
makes
this connection evident, the Brownian motion is just the suitably taken continuum limit 
of the random walk problem. For large $n$, by virtue of the central limit theorem, the results for the 
random walk problem reduce
to those of the Brownian motion. This is an important point 
because in many applications, especially
those in computer science as will be discussed later, one often encounters discrete
random walks as in Eq.~(\ref{rw1}) which are often more difficult to solve than the
continuum Brownian motion. However, since in most applications one is typically interested
in the large time scaling-limit results, one can correctly approximate a discrete random walk sequence by 
the continuum Brownian process and this makes life much simpler. 

\section{Brownian Process as a Path Integral}

The solution of the diffusion equation (\ref{diff3}) can be easily obtained in free space by the Fourier 
transform method. For simplicity, we set $D=1/2$ for the rest of the article. One gets
\begin{equation} 
P(x,t) = \int_{-\infty}^{\infty} dx_0\, G_0(x,t|x_0,0)\, P(x_0,0)
\label{dsol1}
\end{equation}
where $P(x_0,0)$ is the initial condition and the diffusion propagator
\begin{equation}
G_0(x,t|x_0,0)= \frac{1}{\sqrt{2\pi t}}\, \exp\left[-(x-x_0)^2/{2t}\right]
\label{green1}
\end{equation}
denotes the conditional probability that the Brownian particle reaches $x$ at time $t$, starting 
from
$x_0$ at $t=0$. It was M. Kac who first made the important observation~\cite{Kac} that this diffusion
propagator can be interpreted, using Feynman's path integral formalism, as the quantum
propagator of a free particle from time $0$ to time $t$. This is easy to see. Using the
property of the Gaussian noise in Eq.~(\ref{pi0}) and the Langevin equation (\ref{lange1}),
it is clear that the probability of any path $\{x(\tau)\}$ can be written as
\begin{equation}
P\left[\{x(\tau)\}\right]\propto \exp\left[-\frac{1}{2}\,\int_0^{t} \left(\frac{dx}{d\tau}\right)^2\, d\tau\right].
\label{pi1}
\end{equation}
Thus the diffusion propagator, i.e. the probability that a path goes from $x_0$ at $t=0$ to $x$ at $t$
can be written as a sum of over the contributions from all possible paths propagating
from $x_0$ at $\tau=0$ to $x$ at $\tau=t$. This sum is indeed Feynman's path integral~\cite{FH}
\begin{equation}
G_0(x,t|x_0,0) =\int_{x(0)=x_0}^{x(t)=x} {\cal D} x(\tau)\, \exp\left[-\frac{1}{2}\,\int_0^{t} 
\left(\frac{dx}{d\tau}\right)^2\, d\tau \right].
\label{prop0}
\end{equation}
One immediately identifies the term $\frac{1}{2}\,
\left(\frac{dx}{d\tau}\right)^2$ as the classical kinetic energy of a particle of unit mass
and the integral $\frac{1}{2}\,\int_0^{t}
\left(\frac{dx}{d\tau}\right)^2\, d\tau$ as the Lagrangian of a free particle of unit mass.
Following Feynman\cite{FH}, one then identifies the path integral in Eq.~(\ref{prop0}) as
a quantum propagator
\begin{equation}
G_0(x,t|x_0,0)= <x|e^{-{\hat H_0}t}|x_0>
\label{prop1}
\end{equation}
where ${\hat H_0}\equiv -\frac{1}{2}\frac{\partial^2}{\partial x^2}$ is the quantum Hamiltonian
of a free particle (we have set the mass $m=1$ and the Planck's constant $\hbar =1$).
To make the connection complete, the quantum propagator on the r.h.s of Eq.~(\ref{prop1}) can
be easily evaluated by expanding it in the free particle eigenbasis. Noting that $\hat H_0$ has free 
particle eigenfunctions 
$\psi_k(x)= \frac{1}{\sqrt{2\pi}} e^{ikx}$ with eigenvalue $k^2/2$, one gets
\begin{equation}
G_0(x,t|x_0,0)= <x|e^{-{\hat H_0}t}|x_0> =  \int_{-\infty}^{\infty} <x|k><k|x_0> e^{-k^2t/2} dk 
= \frac{1}{2\pi} \int_{-\infty}^{\infty} e^{ik(x-x_0)-k^2t/2}\, dk.
\label{prop2}
\end{equation}
Performing the Gaussian integration, one gets back the classical result in Eq.~(\ref{green1}) that
was obtained by solving the diffusion equation. Thus the two approaches, one by solving a partial 
differential 
equation usually referred to as the Fokker-Planck approach and the other using the path integral method
are completely equivalent.

One may argue that once the basic propagator is known, the Brownian motion is well understood and there is
nothing else interesting left to study! This is simply not true because there are intricate
questions associated with the diffusion process that are often rather nontrivial.
A notable nontrivial example is the calculation of the persistence exponent associated
with a diffusion process~\cite{review}. Consider a diffusive field $\phi(\vec r,t)$ evolving
via the $d$-dimensional diffusion equation
\begin{equation}
\frac{\partial \phi}{\partial t}= \nabla^2 \phi
\label{dpers1}
\end{equation}
starting from the initial condition $\phi(\vec r,0)$ which is a random Gaussian field, uncorrelated in 
space. The solution at time $t$ can be easily found using $d$-dimensional trivial generalization of 
the diffusion propagator in Eq.~(\ref{green1})
\begin{equation}
\phi(\vec r, t) = \frac{1}{(2\pi t)^{d/2}}\,\int d\vec r_0\, \phi(\vec r_0, 0)\, \exp\left[-(\vec r-\vec r_0)^2/{2t}\right].
\label{dpers2}
\end{equation}
Now, suppose that we fix a point $\vec r$ in space and monitor the field $\phi(\vec r,t)$ there as a function of time $t$
and ask: what is the probability $P(t)$ that the field $\phi(\vec r,t)$ at $\vec r$ does not change sign
up to time $t$ starting initially at the random value $\phi(\vec r,0)$? By translational invariance,
$P(t)$ does not depend on the position $\vec r$. This probability $P(t)$ is
called the persistence probability that has generated a lot of interest over the last decade in the 
context of nonequilibrium systems~\cite{review}. For the simple diffusion process in Eq. (\ref{dpers2}),
it is known, both theoretically~\cite{diffusion} and experimentally~\cite{diffexp} that at late times
$t$, the persistence $P(t)$ has a power law tail $P(t)\sim t^{-\theta}$ where the persistence exponent
$\theta$ is nontrivial (even in one dimension!), e.g., $\theta\approx 0.1207 $ in $d=1$, $\theta\approx     
0.1875$ in $d=2$, $\theta\approx 0.2380$ in $d=3$ etc. While this exponent $\theta$ is known
numerically very precisely and also very accurately by approximate analytical methods\cite{diffusion},
an exact calculation of $\theta$ has not yet been achieved and it remains as an outstanding unsolved
problem for the diffusion process~\cite{Watson}. This example thus clarifies that while the knowledge of
the diffusion propagator is necessary, it is by no means sufficient to answer more detailed
history related questions associated with the diffusion process.

Note that in the persistence problem discussed above, the relevant stochastic process at a
fixed point $\vec r$ in space, whose properties one is interested in, is actually
a more complex non-Markovian process~\cite{review} even though it originated from a simple
diffusion equation. In this article, we will stay with our simple Brownian motion
in Eq.~(\ref{lange1}) which is a Markov process and discuss 
some of the nontrivial aspects of this simple Brownian motion.
For example, in many applications of Brownian motion in physics,
finance and computer science, the relevant Brownian process is often constrained. For example, an important
issue is the first-passage property of a Brownian motion~\cite{Chandra,Feller,Redner}, i.e. the 
distribution of the first time that a 
Brownian process crosses the origin? For this, one needs to sample only a subset of all possible Brownian paths
that do not cross the origin up to a certain time. This can be achieved by imposing the constraint of
no crossing on a Brownian path. Apart from the constrained Brownian motion,  some other applications require
a knowledge of the statistical properties of
a Brownian functional up to time $t$, defined as $T_t= \int_0^{t} U\left(x(\tau)\right)\, d\tau$, where 
$U(x)$ is
a specified function. We will provide several examples later and will see that while the
properties of a free Brownian motion are rather simple and are essentially encoded in its propagator
in Eq.~(\ref{green1}), properties of constrained 
Brownian motion
or that of a Brownian functional are often nontrivial to derive and the path integral technique discussed     
above is particularly suitable to address some of these issues.

\subsection{Brownian motion with constraints: first-passage property}

As a simple example of a constrained Brownian motion, we calculate in this subsection the
first-passage probability density $f(x_0,t)$. The quantity $f(x_0,t)dt$ is simply the probability
that a Brownian path, starting at $x_0$ at $t=0$, will cross the origin for the first time between
time $t$ and $t+dt$. Clearly, $f(x_0,t)= -dq(x_0,t)/dt$ where $q(x_0,t)$ is the probability that
the path starting at $x_0$ at $t=0$ does not cross the origin up to $t$.
The probability $q(x_0,t)$ can be easily expressed in terms of
a path integral
\begin{equation}
q(x_0,t) = \int_0^{\infty} dx \int_{x(0)=x_0}^{x(t)=x} 
{\cal D} x(\tau)\, \exp\left[-\frac{1}{2}\,\int_0^{t}
\left(\frac{dx}{d\tau}\right)^2\, d\tau \right]\, \prod_{\tau=0}^{t} \theta\left[x(\tau)\right]
\label{fp1}
\end{equation}
where the paths propagate from the initial position $x(0)=x_0$ to the final position $x$ at time 
$t$ and then we integrate $x$ over only the positive half-space since
the final position $x$ can only be positive. The term
$\prod_{\tau=0}^{t} \theta\left[x(\tau)\right]$ inside the path integral is an indicator function that enforces
the constraint that the path stays above the origin up to $t$. We then identify the path integral in 
Eq.~(\ref{fp1}) 
as an integral over a quantum propagator, 
\begin{equation}
q(x_0,t) = \int_0^{\infty} dx\, G(x,t|x_0,0);\quad\quad\, G(x,t|x_0,0)=<x|e^{-{\hat H_1}t}|x_0>
\label{fp2}
\end{equation}
where the Hamiltonian ${\hat H_1}\equiv -\frac{1}{2}\frac{\partial^2}{\partial x^2} + V(x)$ with
the quantum potential $V(x)=0$ if $x>0$ and $V(x)=\infty$ if $x\le 0$. The infinite potential for $x\le 0$
takes care of the constraint that the path can not cross the origin, i.e. it enforces the condition
$\prod_{\tau=0}^{t} \theta\left[x(\tau)\right]$. The eigenfunction of $\hat H_1$ must vanish at $x=0$, but
for $x>0$ it corresponds to that of a free particle. The correctly normalized eigenfunctions are thus
$\psi_k(x)= \sqrt{\frac{2}{\pi}}\, \sin (kx) $ with $k\ge 0$ with eigenvalues $k^2/2$. The quantum  
propagator
can then be evaluated again by decomposing into the eigenbasis
\begin{equation}
G(x,t|x_0,0)= \frac{2}{\pi}\int_0^{\infty} \sin(k\,x_0)\sin(k\,x) e^{-k^2t/2}\, dk=
\frac{1}{\sqrt{2\pi t}}\left[e^{-(x-x_0)^2/{2t}}-e^{-(x+x_0)^2/{2t}}\right].
\label{fp3}
\end{equation}
Note that this result for the propagator can also be derived alternately by solving the diffusion equation
with an absorbing boundary condition at the origin. The result in Eq.~(\ref{fp3}) then follows by
a simple application of the image method~\cite{Chandra,Redner}.
Integrating over the final position in $x$ one gets from Eq.~(\ref{fp2}) the classical
result~\cite{Feller},
$q(x_0,t)={\rm erf}(x/\sqrt{2t})$
where ${\rm erf}(z)= \frac{2}{\sqrt{\pi}}\,\int_0^{z} e^{-u^2} du$.
The first-passage probability density
is then given by
\begin{equation}
f(x_0,t)= -\frac{d q(x_0,t)}{dt}=\frac{x_0}{\sqrt{2\pi}}\, \frac{e^{-x_0^2/{2t}}}{t^{3/2}}.
\label{fp4}
\end{equation}
For $t\gg x_0^2$, one thus recovers the well known $t^{-3/2}$ decay of the first-passage 
probability~\cite{Chandra,Feller} density.

\subsection{Brownian functionals: Feynman-Kac formula}

In this subsection we will discuss how to calculate the statistical properties of
a Brownian functional defined as
\begin{equation}
T=\int_0^{t} U\left(x(\tau)\right) d\tau
\label{func1}
\end{equation}
where $x(\tau)$ is a Brownian path starting from $x_0$ at $\tau=0$ and propagating 
up to time $\tau=t$ and $U(x)$ is a specified function. 
Clearly $T$ is random variable taking different values for different Brownian paths.
The goal is to calculate its probability distribution $P(T,t|x_0)$. 
The choice of $U(x)$ depends on which quantity we want to calculate.
Brownian functionals appear in a wide range of problems across different fields
ranging from probability theory, finance, data analysis, disordered systems,
and computer science. We consider a few examples below.

\begin{enumerate}

\item In probability theory, an important object of interest is the occupation time, i.e.
the time spent by a Brownian motion above the origin within a time window of size $t$~\cite{Levy}.
Thus the occupation time is simply, $T=\int_0^{t} \theta[x(\tau)]d\tau$. Thus,  
in this problem the function $U(x)=\theta(x)$.

\item For fluctuating $(1+1)$-dimensional interfaces of the Edwards-Wilkinson~\cite{Edwards}
or the Kardar-Parisi-Zhang (KPZ)~\cite{KPZ} varieties, the interface profile in the
stationary state is described by a one dimensional Brownian motion in space~\cite{interreview}.
The fluctuations in the stationary state are captured by the pdf of the spatially
averaged variance of height fluctuations~\cite{Width} in a finite system of size $L$, i.e. 
the pdf of $\sigma^2 = \frac{1}{L}\, \int_0^L h^2(x) dx$ 
where $h(x)$ is the deviation of the height from its spatial average. Since $h(x)$
performs a Brownian motion in space, $\sigma^2$ is a functional of the Brownian
motion as in Eq.~(\ref{func1}) with $U(x) = x^2$.

\item In finance, a typical stock price $S(\tau)$ is sometimes modelled by the exponential
of a Brownian motion, $S(\tau)=e^{-\beta x(\tau)}$, where $\beta$ is a constant. An object that 
often plays a crucial
role is the integrated stock price up to some `target' time $t$, i.e. $T=\int_0^{t} 
e^{-\beta x(\tau)}d\tau$~\cite{Yor}. Thus in this problem $U(x)=e^{-\beta x}$. Interestingly, this 
integrated
stock price has an interesting analogy in a disordered system where a single overdamped particle
moves in a random potential. A popular model is the so called Sinai model~\cite{Sinai}
where the random potential is modelled as the trace of a random walker in space.
Interpreting the 
time $\tau$ as the spatial distance, $x(\tau)$ is then the potential energy of the
particle and $e^{-\beta x(\tau)}$ is just the Boltzmann factor. The total
time $t$ is just the size of a linear box in which the particle is moving.
Thus $T=\int_0^{t}
e^{-\beta x(\tau)}d\tau$ is just the partition function of the particle
in a random potential~\cite{CMY}. 
In addition, the exponential of a Brownian motion also appears in the expression
for the Wigner time delay in one dimensional quantum scattering process by
a random potential~\cite{CT}.

\item In simple models describing the stochastic behavior of daily temperature records,
one assumes that the daily temperature deviation from its average is a simple
Brownian motion $x(\tau)$ in a harmonic potential (the Ornstein-Uhlenbeck process). Then
the relevant quantity whose statistical properties are of interest is the
so called `heating degree days' (HDD) defined as $T=\int_0^{t} x(\tau)\, \theta(x(\tau))\,d\tau$
that measures the integrated excess temperature up to time $t$~\cite{MB}. Thus in this example,
the function $U(x)= x\theta(x)$.   

\item Another quantity, first studied in the context of economics~\cite{CifaReg} and later extensively by 
probabilists~\cite{Shepp} is the
total area (unsigned) under a Brownian motion, i.e. $T=\int_0^{t} |x(\tau)|\,d\tau$,
Thus in this example, $U(x)=|x|$. The same functional was also studied by physicists
in the context of electron-electron and phase coherence in one dimensional
weakly disordered quantum wire\cite{AKK}.

\end{enumerate}

We will mention several other examples as we go along. Note that in all the examples
mentioned above the function $U(x)$ is such that the random variable $T$ has only
positive support. Henceforth we will assume that. For a given such function 
$U(x)$, how does one calculate the pdf of $T$? 
It was Kac who, using the path integral techniques developed by Feynman in his
Ph.D  thesis, first devised a way of computing the pdf $P(T,t|x_0)$ of a Brownian
functional~\cite{Kac} that led to the famous Feynman-Kac formula. 
We summarize below Kac's formalism.

\vspace{0.4cm}

\noindent {\bf Feynman-Kac formula:}  Since $T$ has only positive 
support, a natural step
is to introduce the Laplace transform of the pdf $P(T,t|x_0)$,
\begin{equation}
Q(x_0,t)= \int_0^{\infty} e^{-p\, T} P(T,t|x_0) dT
=E_{x_0}\left[e^{-p\, \int_0^{t} U(x(\tau))d\tau}\right]
\label{expec1}
\end{equation}
where the r.h.s is an expectation over all possible Brownian paths $\{x(\tau)\}$ that start at $x_0$
at $\tau=0$ and propagate up to time $\tau=t$. We have, for notational simplicity, suppressed the $p$ dependence  
of $Q(x_0,t)$. Using the measure of the Brownian path in Eq.~(\ref{pi1}), one can then express the
expectation on the r.h.s of Eq.~(\ref{expec1}) as a path integral
\begin{eqnarray}
Q(x_0,t)=E_{x_0}\left[e^{-p\, \int_0^{t} U(x(\tau))d\tau}\right]&=& \int_{-\infty}^{\infty} dx 
\int_{x(0)=x_0}^{x(t)=x} {\cal D} x(\tau)\, 
\exp\left[-\int_0^t d\tau \left[\frac{1}{2}
\left(\frac{dx}{d\tau}\right)^2\, + p\, U(x(\tau)\right]\right] \label{expec2} \\
&=& \int_{-\infty}^{\infty} dx <x| e^{-{\hat H}t}|x_0>
\label{expec3}
\end{eqnarray}
where the quantum Hamiltonian ${\hat H}\equiv -\frac{1}{2}\frac{\partial^2}{\partial x^2} + p U(x)$
corresponds to the Shr\"odinger operator with a potential $p U(x)$. Note that in Eq.~(\ref{expec2}) 
all
paths propagate from $x(0)=x_0$ to $x(t)=x$ in time $t$ and then we have 
integrated over the final position $x$. The quantum propagator $G(x,t|x_0)=<x|e^{-{\hat H}t}|x_0>$
satisfies a Shr\"odinger like equation 
\begin{equation}
\frac{\partial G}{\partial t}= \frac{1}{2} \frac{\partial^2 G}{\partial x^2} - p\,U(x)\, G
\label{ffp1}
\end{equation}
which can be easily established by differentiating $G(x,t|x_0)=<x|e^{-{\hat H}t}|x_0>$ with respect 
to $t$
and using the explicit representation of the operator $\hat H$. The initial condition
is simply, $G(x,0|x_0)= \delta(x-x_0)$. Thus the scheme of Kac 
involves
three steps: (i) solve the partial diferential equation (\ref{ffp1}) to get $G(x,t|x_0)$
(ii) integrate $G(x,t|x_0)$ over the final position $x$ as in Eq.~(\ref{expec3}) to obtain
the Laplace transform $Q(x_0,t)$ and (iii) invert the Laplace transform in Eq.~(\ref{expec1})
to obtain the pdf $P(T,t|x_0)$. The equations (\ref{expec1}), (\ref{expec3}) and
(\ref{ffp1}) are collectively known as the celebrated Feynman-Kac formula.

\vspace{0.4cm}

\noindent {\bf A shorter backward Fokker-Planck approach:} An alternative and somewhat shorter 
approach would be to write down a partial differential
equation for $Q(x_0,t)$ in Eq.~(\ref{expec3}) directly. An elementary exercise yields
\begin{equation}
\frac{\partial Q}{\partial t}= \frac{1}{2} \frac{\partial^2 Q}{\partial x_0^2}-p\,U(x_0)\, Q
\label{bfp1}
\end{equation}
where note that the spatial derivatives are with respect to the {\em initial} position $x_0$.
This is thus a `backward' Fokker-Planck approach as opposed to the `forward' Fokker-Planck
equation satisfied by $G$ in Eq.~(\ref{ffp1}) of Kac where the spatial derivatives are
with respect to the {\em final} position of the particle. Basically we have 
reduced the
additional step (ii) of integrating over the final position in Kac's derivation. 
The solution $Q(x_0,t)$ of Eq.~(\ref{bfp1}) must satisfy the initial condition
$Q(x_0,0)=1$ that follows directly from the definition in Eq.~(\ref{expec1}).
To solve Eq.~(\ref{bfp1}), it is useful to    
take a further Laplace transform of Eq.~(\ref{bfp1}) with respect to $t$,
${\tilde Q}(x_0,\alpha)= \int_0^{\infty} Q(x_0,t) e^{-\alpha t} dt$. 
Using the initial
condition $Q(x_0,0)=1$, one arrives at an ordinary second order differential equation
\begin{equation}
\frac{1}{2}\frac{d^2 {\tilde Q}}{dx_0^2}-[\alpha+p\, U(x_0)]{\tilde Q} = -1 
\label{bfp2}
\end{equation}
which needs to be solved subject to the appropriate boundary conditions that depend
on the behavior of the function $U(x)$ at large $x$. Given that $T=\int_0^{t} 
U\left(x(\tau)\right)\,d\tau$
has positive support, there are two typical representative asymptotic behaviors
of $U(x)$:
\begin{enumerate}

\item If the function $U(x)$ approaches a constant value at large $x$, i.e.
$U(x)\to c_{\pm} $ as $x\to \pm \infty$, then it is easy to argue (for an example, see below)
that $Q(x_0\to \pm \infty, \alpha) = 1/[p\,c_{\pm} + \alpha]$. In this case, the
underlying quantum Hamiltonian ${\hat H}\equiv -\frac{1}{2}\frac{\partial^2}{\partial x^2} + p 
U(x)$ has scattering states in its spectrum, in addition to possible bound states.

\item If the function $U(x)\to \infty$ as $x\to \pm \infty$, then $Q(x_0\to \pm \infty, \alpha)=0$.
In this case the underlying quantum Hamiltonian ${\hat H}$ has only bound states and hence a
discrete spectrum.

\end{enumerate}
Thus, in principle,
knowing the solution ${\tilde Q}(x_0,\alpha)$ of Eq.~(\ref{bfp2}), the original pdf $P(T,t|x_0)$ 
can be
obtained by inverting the double Laplace transform
\begin{equation}
{\tilde Q}(x_0,\alpha)= \int_0^{\infty} dt\, e^{-\alpha t}\int_0^{\infty} dT\, e^{-pT}\, P(T,t|x_0).
\label{dlt0}
\end{equation} 
Below we provide an example where all these steps can be carried out explicitly to obtain
an exact closed form expression for the pdf $P(T,t|x_0)$.

\subsection{ A simple illustration: L\'evy's arcsine law for the distribution of the occupation time}  

As an illustration of the method outlined in the previous subsection, let us calculate the distribution of the 
occupation time $T=\int_0^{t} \theta[x(\tau)]d\tau$. This distribution was first computed by
L\'evy using probabilistic methods~\cite{Levy}. Later Kac derived it using Feynman-Kac formalism discussed
above~\cite{Kac}. We present here a derivation based on the backward Fokker-Planck approach outlined above. 

Substituting $U(x_0)=\theta(x_0)$ in Eq.~(\ref{bfp2})
we solve the differential equation separately for $x_0>0$ and $x_0<0$ and then match 
the solution at $x_0=0$ by demanding the continuity of the solution and that of its first derivative.
In addition, we use the boundary conditions ${\tilde Q}(x_0\to \infty, \alpha)=1/(\alpha+p)$
and ${\tilde Q}(x_0\to -\infty, \alpha)= 1/\alpha$. They follow from the observations:
\begin{enumerate}

\item If the starting point $x_0\to \infty$, the particle will stay on the positive side
for all finite $t$ implying $T=\int_0^t \theta(x(\tau))\,d\tau=t$ and hence 
$Q(x_0\to \infty, t)= E[e^{-pT}]=e^{-pt}$
and its Laplace transform ${\tilde Q}(x_0\to \infty, \alpha)=\int_0^{\infty} e^{-(\alpha+p)\,t}dt
=1/(\alpha+p)$.

\item If the starting point $x_0\to -\infty$, the particle stays on the negative side up to any
finite $t$ implying $T=\int_0^t \theta(x(\tau))\, d\tau=0$ and hence 
$Q(x_0\to -\infty, t)=E[e^{-pT}]=1$ and its Laplace transform ${\tilde Q}(x_0\to -\infty, \alpha)=
\int_0^{\infty} e^{-\alpha t} dt= 1/\alpha$.

\end{enumerate}

Using these boundary and matching conditions, one obtains an explicit solution
\begin{eqnarray}
{\tilde Q} (x_0, \alpha)&=& \frac{1}{(\alpha+p)}\left[1+ \frac{(\sqrt{\alpha+p}-\sqrt{\alpha})}{\sqrt{\alpha}}\, 
e^{-\sqrt{2(\alpha+p)}\, x_0}\right]\quad\quad {\rm for}\,\, x_0>0 \label{pos} \\
&=& \frac{1}{\alpha}\left[1+ \frac{(\sqrt{\alpha}-\sqrt{\alpha+p})}{\sqrt{\alpha+p}}\,
e^{\sqrt{2\,\alpha}\, x_0}\right]\quad\quad\quad\quad\quad\quad {\rm for}\,\, x_0<0 .
\label{neg}
\end{eqnarray}
The solution is simpler if the particle starts at the origin $x_0=0$. Then one gets from above
\begin{equation}
{\tilde Q}(0, \alpha)= \frac{1}{\sqrt{\alpha(\alpha+p)}}.
\label{dlt1}
\end{equation}
Inverting the Laplace transform, first with respect to $p$ and then with respect to $\alpha$, one obtains 
the pdf of the occupation time for all $0\le T\le t$
\begin{equation}
P(T, t|x_0=0) = \frac{1}{\pi}\, \frac{1}{\sqrt{T(t-T)}}.
\label{occup1}
\end{equation}
In particular, the cumulative distribution 
\begin{equation}
\int_0^{T} P(T', t|x_0=0) dT'= \frac{2}{\pi} \arcsin\left(\sqrt{\frac{T}{t}}\right)
\label{levy1}
\end{equation}
is known as the famous arcsine law of L\'evy~\cite{Levy}.

The result in Eq.~(\ref{occup1}) is interesting and somewhat counterintuitive. The probability
density peaks at the two end points $T=0$ and $T=t$ and has a minimum at $T=1/2$ which is also
the average occupation time. Normally one would expect that any `typical' path would spend
roughly half the time $t/2$ on the positive side and the other half on the negative side.
If that was the case, one would have a peak of the occupation time distribution at
the average value $t/2$. The actual result is exactly the opposite--one has a minimum
at $T=t/2$! This means that a typical 
path, starting at the origin, tends to stay
either entirely on the positive side (explaining the peak at $T=t$) or entirely
on the negative side (explaining the peak at $T=0$). 
In other words, a typical Brownian path is `stiff' and reluctant to cross the origin.
This property that `the typical is not the same as the average' is one of the 
hidden surprises of Einstein's Brownian motion.

The concept of the occupation time and related quantities have been studied
by probabilists for a long time~\cite{Lamperti}. Recently they have played important  
roles in physics as well, for example in
understanding
the dynamics out of equilibrium in coarsening systems~\cite{occup0}, ergodicity
properties in anomalously diffusive processses~\cite{DM},
in renewal processes~\cite{GL}, 
in models related to spin glasses~\cite{MD1}, in understanding
certain aspects of transport properties in disordered sysyems~\cite{occup1} and
also in simple models of blinking quantum dots~\cite{BB}.

\section{Area under a Brownian Excursion: Applications in Physics and Computer Science}

In this section we consider an example where, by applying the path integral
method outlined in the previous section, one can compute exactly the
distribution of a functional of a Brownian process that is also 
constrained to stay positive over a fixed time interval $[0,t]$. 
A Brownian motion $x(\tau)$ in an interval $0\le \tau\le t$, that starts and ends at the origin 
$x(0)=x(t)=0$ but is conditioned to stay positive in between, is called a Brownian
excursion. The area under the excursion, $A=\int_0^T x(\tau) d\tau$, is clearly a random
variable taking a different value for each realization of the excursion. A natural question that 
the mathematicians have studied quite extensively~\cite{Darling,Louchard,Takacs,FPV,FL} over the past two 
decades is: what is the pdf
$P(A,t)$ of the area
under a Brownian excursion over the interval $[0,t]$? Since the typical lateral displacement of 
the excursion
at time $\tau$ scales as $\sqrt{\tau}$, it follows that the area over the interval $[0,t]$ will scale as
$t^{3/2}$ and hence its pdf must have a scaling form,
$P(A,t)= t^{-3/2}
f\left(A/t^{3/2}
\right)$. The normalization condition $\int_0^{\infty} P(A,t)dA=1$ demands a prefactor 
$t^{-3/2}$
and also the conditions: $f(x)\ge 0$ for all $x$ and  $\int_0^{\infty} f(x)dx=1$. One then 
interprets the
scaling function $f(x)$ as the
distribution of the
area under the Brownian excursion $x(u)$ over a {\em unit} interval $u\in [0,1]$.
The function $f(x)$, or rather its Laplace transform, was first computed analytically
by Darling~\cite{Darling} and independently by Louchard~\cite{Louchard},
\begin{equation}
{\tilde f} (s)= \int_0^{\infty} f(x) e^{-sx} dx= s\sqrt{2\pi}\sum_{k=1}^{\infty} e^{-\alpha_k 
s^{2/3}
2^{-1/3}},
\label{airy1}
\end{equation}
where $\alpha_k$'s are the magnitudes of the zeros of the standard Airy function ${\rm Ai}(z)$.
The Airy function ${\rm Ai}(z)$ has discrete zeros on the
negative real axis at e.g. $z=-2.3381$, $z=-4.0879$, $z=-5.5205$ etc. Thus, $\alpha_1=2.3381\dots$, 
$\alpha_2=4.0879\dots$, $\alpha_3=5.5205\dots$ 
etc. 
Since the expression of $f(x)$ involves the zeros of Airy function, the function $f(x)$ has been 
named the 
Airy distribution function~\cite{FPV}, which should not be confused with the Airy function ${\rm Ai}(x)$ 
itself.
Even though Eq.~(\ref{airy1}) provides a formally exact expression of the Laplace transform, 
it turns out that
the calculation of the moments $M_n= \int_0^{\infty} x^n f(x)dx$ is highly nontrivial and they 
can be
determined only recursively~\cite{Takacs} (see Section II). Takacs was able to formally 
invert the Laplace transform
in Eq.~(\ref{airy1}) to obtain~\cite{Takacs},
\begin{equation}
f(x)= {\frac{2\sqrt{6}}{x^{10/3}}}\sum_{k=1}^{\infty} e^{-b_k/x^2} b_k^{2/3}
U(-5/6, 4/3, b_k/x^2),
\label{fx1}
\end{equation}
where $b_k= 2\alpha_k^3/{27}$ and $U(a,b,z)$ is the confluent hypergeometric function~\cite{AS}.
The function $f(x)$ has the following asymptotic tails~\cite{Takacs,CSY},
\begin{eqnarray}
f(x) &\sim & x^{-5}\, e^{- 2\alpha_1^3/{27 x^2}} \quad {\rm as}\quad x\to 0 \nonumber \\
f(x) &\sim & e^{-6 x^2} \quad\quad\quad\quad\quad {\rm as}\quad x\to \infty.
\label{asymfx}
\end{eqnarray}
So, why would anyone care about such a complicated function? The reason behind the sustained 
interest and study~\cite{Takacs,FPV,FL,PW} of this function $f(x)$ seems to be the fact that it keeps
resurfacing in a number of seemingly unrelated problems, in computer science, graph 
theory, two dimensional growth problems and more recently in fluctuating one dimensional
interfaces. We discuss below some of these applications.

\begin{figure}
\includegraphics[height=10.0cm,width=10.0cm,angle=0]{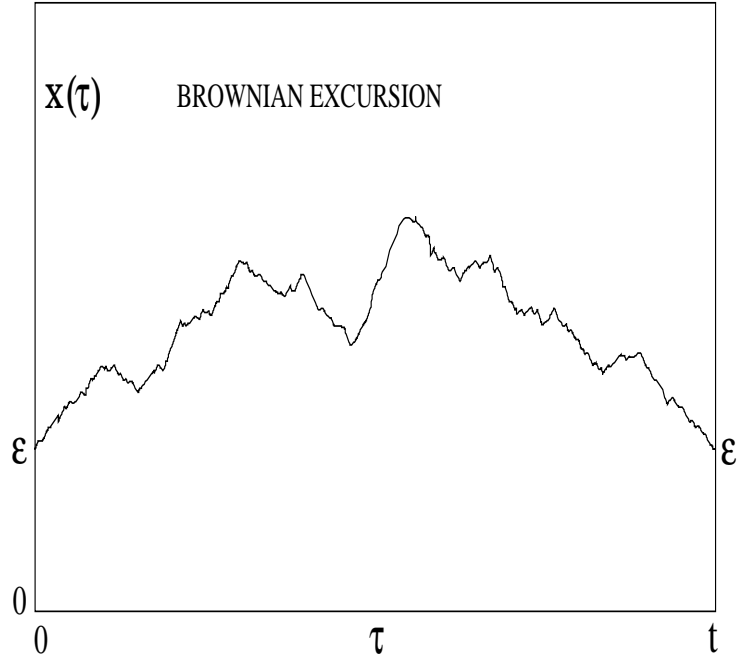}
\caption{\label{fig:excur1} A Brownian excursion over the time interval $0\le \tau \le t$ starting 
at $x(0)=\epsilon$ and ending at $x(t)=\epsilon$ and staying positive in between.}
\end{figure}

The result in Eq.~(\ref{airy1}) was originally derived using probabilistic 
methods~\cite{Darling,Louchard}. 
A more direct physical derivation using the path integral method was provided more recently~\cite{MC2}, 
which we outline below. Following the discussion in the previous section, our interest 
here is in the functional $T=A=\int_0^{t} x(\tau)d\tau$.
However, we also need to impose the constraint that the path stays positive between $0$ and $t$,
i.e. we have to insert a factor $\prod_{\tau} \theta[x(\tau)]$ in the path integral.
However, one needs to be a bit careful in implementing this constraint. Note that
the path starts at the origin, i.e. $x(0)=0$. But if we take a continuous time Brownian path 
that starts at the origin, it immediately recrosses the origin many times and hence
it is impossible to restrict a Brownian path to be positive over an interval if it starts at 
the origin. One can circumvent this problem by introducing a small cut-off $\epsilon$, i.e.
we consider all paths that start at $x(0)=\epsilon$ and end at $x(t)=\epsilon$ and stays
positive in between (see Fig.~ (\ref{fig:excur1})). We then first derive the pdf 
$P(A,t,\epsilon)$
and then take the limit $\epsilon\to 0$ eventually. 

Following the method in the previous 
section, the Laplace transform of the pdf is now given by
\begin{equation}
Q(\epsilon,t) = E_{\epsilon}\left[e^{-p \int_0^{t} x(\tau) d\tau} \right]= 
\frac{1}{Z_E} \int_{x(0)=\epsilon}^{x(t)=\epsilon} 
{\cal D}x(\tau)\,e^{-\int_0^{t}d\tau\, \left[\frac{1}{2}(dx/{d\tau})^2 + p 
x(\tau)\right]}\,\prod_{\tau=0}^{t}
\theta\left[x(\tau)\right].
\label{exarea1}
\end{equation}
where $Z_E$ is a normalization constant
\begin{equation}
Z_E= \int_{x(0)=\epsilon}^{x(t)=\epsilon} {\cal D}x(\tau)\,
e^{-\frac{1}{2}\int_0^{t}d\tau\, (dx/{d\tau})^2}\, \prod_{\tau=0}^{t}
\theta\left[x(\tau)\right]
\label{expart1}
\end{equation}
that is just the partition function of the Brownian excursion. 

Clearly, $Z_E=<\epsilon|e^{-{\hat 
H_1}t}|\epsilon>$ where ${\hat H_1}\equiv -\frac{1}{2} \frac{d^2}{dx^2} +V(x)$,
with the potential $V(x)=0$ for $x>0$ and $V(x)=\infty$ for $x\le 0$. We have already evaluated 
this in Section III-A in Eq.~(\ref{fp3}). Putting $x=x_0=\epsilon$ in Eq.~(\ref{fp3}) we get
$Z_E= G(\epsilon,t\,|\epsilon,0)= (1-e^{-2\epsilon^2 t})/\sqrt{2\pi t}$. The path integral in the numerator
in Eq.~(\ref{exarea1}) is simply the propagator 
$<\epsilon|e^{-{\hat H} t}|\epsilon>$ where the
Hamiltonian $\hat H\equiv  -\frac{1}{2} \frac{d^2}{dx^2} + pU(x)$ with a triangular
potential $U(x) = x$ for $x>0$ and $U(x)=\infty$ for $x\le 0$. The Hamiltonian
$\hat H$ has only bound states and discrete eigenvalues. Its eigenfunctions are
simply shifted Airy functions and eigenvalues are given by the negative of the zeros of the
Airy function.  
Expanding the propagator into its eigenbasis and
finally taking the $\epsilon \to 0$ limit (for details see Ref. ~\cite{MC2}), one derives the 
result
\begin{equation}
Q(0,t) =\int_0^{\infty} P(A,t)\, e^{-p\,A}\, dA= \sqrt{2\pi}\, (p t^{3/2})\, \sum_{k=1}^{\infty} e^{-2^{-1/3}\, 
\alpha_k\, (p t^{3/2})^{2/3}}
\label{area2}
\end{equation}
where $\alpha_k$'s are the negative of the zeros of the Airy function. The result in 
Eq.~(\ref{area2}) indicates that its inverse Laplace transform has the scaling form, $P(A,t)= 
t^{-3/2}\, f(A t^{-3/2})$ where the Laplace transform of the scaling function $f(x)$ is given 
in Eq.~(\ref{airy1}).

\vspace{0.4cm}

\noindent {\bf Applications of the Airy Distribution Function: }
The Airy distribution function in Eq.~(\ref{fx1}) has appeared in a number of applications
ranging from computer science and graph theory to physics. Below we mention some of these
applications.

\vspace{0.4cm}

\noindent {\bf 1. Cost function in data storage:} One of the simplest algorithms for data storage
in a linear table 
is called the linear probing with hashing (LPH) algorithm. It was originally introduced by D. 
Knuth~\cite{Knuth}
and has been the object of intense study in computer science due to its simplicity, efficiency
and general applicability~\cite{FPV}. 
Recently it was shown~\cite{droppush} that the LPH algorithm gives rise to a correlated drop-push percolation model
in one dimension that belongs to a different universality class compared to the ordinary
site percolation model.
Knuth, a pioneer in the analysis of algorithms,  has indicated that 
this problem has had a strong influence on his scientific career~\cite{FPV}. The LPH algorithm is described 
as follows: Consider $M$ items 
$x_1$, $x_2$,
$\dots$, $x_M$ to be placed sequentially into a linear table with $L$ cells labelled
$1$, $2$, $\dots$, $L$ where $L\ge M$.
Initially all cells are empty and each cell can contain at most one item. For each item  
$x_i$, a hash address $h_i\in \{1,2,\dots,L\}$ is assigned, i.e. the 
label $h_i$ denotes the address of the cell to which $x_i$ should go. Usually the
hash address $h_i$ is chosen randomly from the set $\{1,2,\dots,L\}$. 
The item $x_i$ is inserted at its hash address $h_i$ provided the cell labelled $h_i$
is empty. If it is already occupied, one tries cells $h_i+1$, $h_i+2$, etc. until an empty cell is found (the 
locations of the cells are interpreted modulo $L$) where the item $x_i$ is finally inserted.
In the language of statistical physics, this is like a drop-push model. One starts with
an empty periodic lattice. A site is chosen at random and one attempts to drop a particle there.
If the target site is empty, the incoming particle occupies it and one starts the process with a
new particle. If the target site is occupied, then the particle keeps hopping to the right until it
finds an empty site which it then occupies and then one starts with a new particle and so on. 

\begin{figure}
\includegraphics[height=10.0cm,width=10.0cm,angle=0]{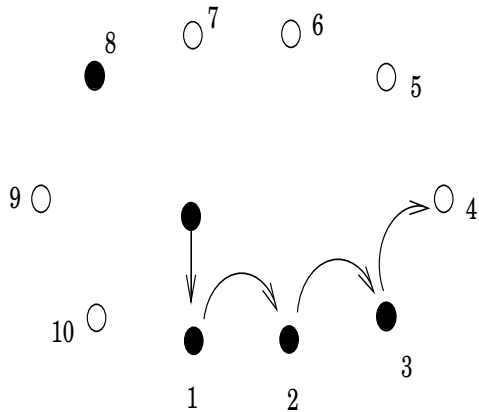}
\caption{\label{fig:lph}
The LPH algorithm for a table of $10$ sites. The figure shows an incoming item
which chose randomly the site $1$ to drop, but since site $1$ is already occupied, the incoming item
keeps hopping to the right until it finds an empty cell at location $4$ to which it gets absorbed.}
\end{figure}

From the computer science point of view, the object of interest is the cost function $C(M,L)$
defined as the total number of unsuccessful probes encountered in inserting the $M$ items 
into a table of size $L$. In particular, the total cost $C=C(L,L)$ in filling up the table
is an important measure of the efficiency of the algorithm. The
cost $C$ is clearly a random variable, i.e. it has different value for different histories
of filling up the table. A central question is: What is its pdf
$P(C,L)$? It has been shown rigorously by combinatorial methods~\cite{FPV} that $P(C,L)$ has a
scaling form for large $L$, $P(C,L) \simeq L^{-3/2} f(C L^{-3/2})$ where the scaling function $f(x)$
is precisely the Airy distribution function in Eq.~(\ref{fx1}) that describes the distribution
of area under a Brownian excursion. To understand the connection between the two problems,
consider any given history of the process where the table, starting initially with all sites
empty, gets eventually filled up.
We define a stochastic quantity $X_i$ that measures the total number of attempts
at site $i$ till the end of the process in any given history. Clearly $X_i\ge 1$ and out of $X_i$ attempts
at site $i$, only one of the attempts (the first one) has been successful in filling up the site,
the rest $(X_i-1)$ of them had been unsuccessful. Thus, the total cost is $C= \sum_{i=1}^L (X_i-1)$.   
Now, the site $(i-1)$ has been attempted $X_{i-1}$ times, out of which only the first one was
successful and the rest $(X_{i-1}-1)$ attempts resulted in pushing the particle to
the right neighbour $i$ and thus each of these unsuccessful attempts at $(i-1)$ result
in an attempt at site $i$. Thus, one can write a recursion relation
\begin{equation}
X_i = X_{i-1}-1 + \xi_i
\label{dp1}
\end{equation}
where $\xi_i$ is a random variable that counts the number of direct attempts (not coming from site
$(i-1)$) at site $i$. Thus ${\rm Prob}(\xi=k)= {\rm Prob}(\rm {the\,\, site\,\, i\,\, is\,\, chosen\,\, for\,\, 
direct\,\, hit\,\,
k\,\, times\,\, out\,\, of\,\, a\,\, total\,\, L\,\, trials})= {L \choose k} (1/L)^k (1-1/L)^{L-k} $, since for random 
hashing, the 
probability 
that site $i$ is chosen, out of $L$ sites, is simply $1/L$. Clearly the noise $\xi$ has a mean value,
$<\xi>= 1$. If we now define $x_i= X_i-1$, then $x_i$'s satisfy
\begin{equation}
x_i= x_{i-1} +\eta_i
\label{dp2}
\end{equation}
where $\eta_i= \xi_i-1$ is a noise, independent from site to site, and for each site $i$, it is
chosen from a binomial distribution. Note that $<\eta_i>=<\xi_i>-1=0$. Thus, $x_i$'s clearly
represent a random walk in space from $0$ to $L$ with periodic boundary conditions.
Moreover, since $X_i\ge 
1$, we have $x_i\ge 0$, indicating that
it is a discrete version of a Brownian excursion and the total cost 
$C= \sum_{i=1}^L (X_i-1)=\sum_{i=1}^L x_i$
is just the area under the Brownian excursion. For large number of steps $L$, the discrete and the continuum version 
share the same probability distribution, thus proving that the probability distribution of the
total cost in LPH algorithm
is precisely the same as that of the area under a Brownian excursion.   

\vspace{0.4cm}

\begin{figure}
\includegraphics[height=10.0cm,width=10.0cm,angle=0]{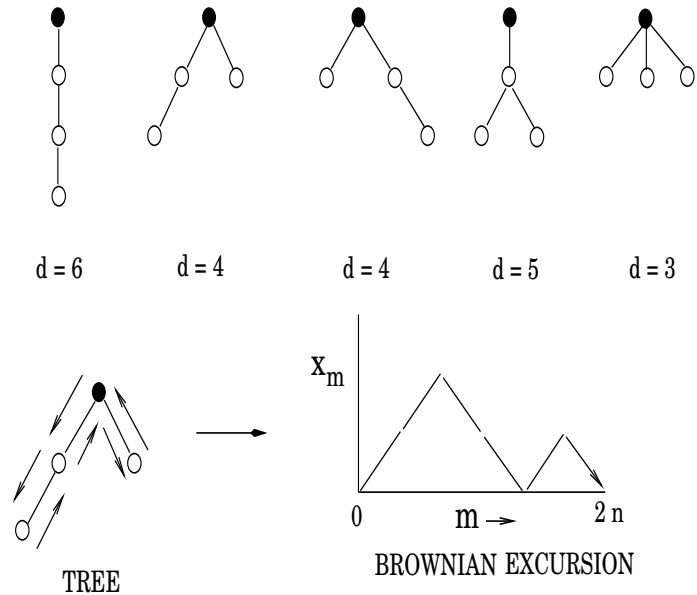}
\caption{\label{fig:tree}
The $5$ possible rooted planar tree with $(3+1)$ vertices. Each configuration has an
associated total internal path length $d$ listed below the configuration. Any given tree
configuration, say the second one in the figure, has a one to one correspondence to a
Dyck path, i.e. a configuration of a Brownian excursion (discrete time random walk).}
\end{figure}

\noindent {\bf 2. Internal path lengths on rooted planar trees:} Rooted planar trees are 
important combinatorial 
objects in graph theory and computer science~\cite{Harary}. Examples of rooted planar trees with $n+1=4$ 
vertices are shown in Fig.~(\ref{fig:tree}). There are 
in general $C_{n+1}= \frac{1}{n+1}{2n\choose n}$ number of possible rooted 
planar tree
configurations with $(n+1)$ vertices. For example, $C_1=1$, $C_2=1$, $C_3=2$, $C_4=5$, $C_6=14$ etc.--these are
the Catalan numbers. An important quantity of interest is the total internal path length $d$ of a tree 
which
is simply the sum of the distances of all the $n$ vertices from the root, $d= \sum_{i=1}^n d_i$,
$d_i$ being the distance of the $i$-th vertex from the root. Each tree configuration has a particular
value of $d$, e.g. in Fig.~(\ref{fig:tree}) the $5$ different configurations have values $d=6$, 
$d=4$, 
$d=4$, $d=5$
and $d=3$ respectively. Suppose that all $C_{n+1}$ configurations of trees for a fixed $n$ are sampled with
equal probability: what is the probability density $P(d,n)$ of the internal path length $d$?
This problem again can be mapped~\cite{Takacs} to the problem of the area under a Brownian excursion as 
shown in Fig.~(\ref{fig:tree}).
Starting from the root of a planar tree with $(n+1)$ vertices, suppose one traverses the vertices of a tree as shown 
by the arrows in Fig.~(\ref{fig:tree}),
ending at the root.
We think of this route as the path of a random walker in one dimension.
For each arrow pointing away from the root on the tree, we draw a step of the random walker with an upward slope.
Similarly, for each arrow pointing to the root on the tree, we draw a step of the random walker with a
downward slope. Since on the tree, one comes back to the root, it is evident by construction that the
corresponding configuration of the random walker $x_m$ is an excursion (i.e. it never goes to the negative side of 
the origin) that starts at the origin and ends up at the origin after $2n$ steps, $x_0=0$ and $x_{2n}=0$.
Such excursions of a discrete random walk are called Dyck paths. Now, the total internal path 
length $d$ of any tree 
configuration is
simply related to the total `area' under a Dyck path via, $2 d= \sum_{m=1}^{2n} x_{m}+n$, as can be easily 
verified.
Now, for large $n$, Dyck paths essentially becomes Brownian excursions and the object $\sum_{m=1}^{2n} x_m$           
is simply the area $A_{2n}$ under a Brownian excursion over the time interval $[0,2n]$. Since $A_{2n}\sim 
(2n)^{3/2}$ for large $n$, it follows that $d\simeq A_{2n}/2$. Therefore, its probability density
$P(d,n)$ has a scaling form, $P(d,n) = \frac{1}{\sqrt{2} n^{3/2}}\,f(d/{{\sqrt 2}n^{3/2}})$ where
$f(x)$ is precisely the Airy distribution function in Eq.~(\ref{fx1}).

\vspace{0.4cm}

\noindent {\bf 3. Maximal relative height distribution for fluctuating interfaces:} 
Fluctuating interfaces have been 
widely studied over the last two decades as they appear in a variety of
physical systems such as growing crystals, molecular beam epitaxy, fluctuating steps on metals and growing
bacterial colonies~\cite{interreview}. The most well studied model of a fluctuating $(1+1)$-dimensional
surfaces is the so called Kardar-Parisi-Zhang (KPZ) equation~\cite{KPZ} that describes the time evolution 
of the height $H(x,t)$ of an interface growing over a linear substrate of size $L$ via the stochastic partial 
differential equation
\begin{equation}
{\frac{\partial H(x,t)}{\partial t}}= {\frac{\partial^2 H(x,t)}{\partial x^2}}+ \lambda 
{\left( { \frac{\partial H(x,t)}{\partial x} } \right) }^2 + \eta(x,t),
\label{kpz1}
\end{equation}
where $\eta(x,t)$ is a Gaussian white noise with zero mean and a correlator, $\langle
\eta(x,t)\eta(x't')\rangle= 2 \delta(x-x')\delta(t-t')$. If the parameter $\lambda=0$, the equation
becomes linear and is known as the Edwards-Wilkinson equation~\cite{Edwards}. 
We consider the general case when $\lambda\ge 0$.
The height is usually measured
relative to the spatially averaged height, i.e. $h(x,t) = H(x,t) - \int_0^{L} H(x',t)dx'/L$.
The joint probability distribution of the relative height field $P(\{h\},t)$ becomes 
time-independent as $t\to \infty$ in a finite system of size $L$. An important quantity
that has created some interests recently~\cite{RCPS,MC1,GK} is the pdf of the
maximal relative height (MRH) in the stationary state, i.e. $P(h_m, L)$ where
\begin{equation}
h_m = \lim_{t\to \infty} {\rm max}_x\,\left[ \{h(x,t)\}, 0\le x\le L\right].
\label{maxh1}
\end{equation}
This is an important physical quantity that measures the extreme fluctuations of
the interface heights. Note that in this system 
the height variables are strongly correlated in the stationary state. While the theory of extremes of a
set of uncorrelated (or weakly correlated) random variables is well established~\cite{Extreme}, not much is 
known
about the distribution of extremes of a set of strongly correlated random variables. Analytical results
for such strongly correlated variables would thus be welcome from the general
theoretical perspective and 
the system of fluctuating interfaces provides exactly the opportunity to study the 
extreme distribution analytically in a strongly correlated system.
This problem of finding the MRH distribution was recently mapped~\cite{MC1,MC2} again to the problem of
the area under a Brownian excursion using the path integral method outlined in section-III
and it was shown that for periodic boundary conditions, $P(h_m,L) = L^{-1/2} f(h_m/\sqrt{L})$
where $f(x)$ is again the Airy distribution function in Eq.~(\ref{fx1}). Interestingly, the
distribution does not depend explicitly on $\lambda$. 
This is thus one of the rare examples where one can calculate analytically the distribution
of the extreme of a set of strongly correlated random variables~\cite{MC1,MC2}. 

\vspace{0.4cm}

\noindent {\bf 4. Other applications:} Apart from the three examples mentioned above, the Airy distribution function and 
its moments
also appear in a number of other problems. For example, The generating
function for the number of inversions in trees involves the Airy distribution function $f(x)$~\cite{MR}.
Also, the moments $M_n$'s of the function $f(x)$ appear in the enumeration of the connected components in
a random graph~\cite{Wright}. Recently, it has been conjectured and subsequently tested numerically
that the asymptotic pdf of the area of two dimensional self-avoiding polygons is also given by the
Airy distribution function $f(x)$~\cite{RGJ}. Besides, numerical evidence suggests that the area enclosed
by the outer boundary of planar random
loops is also distributed according to the Airy distribution function $f(x)$~\cite{RGJ}.  

\section{First-passage Brownian Functional}

So far we have studied the pdf of a Brownian functional over a fixed time interval $[0,t]$.
In this section, we show how to compute the pdf of a Brownian functional over the time
interval $[0,t_f]$ where $t_f$ is the first-passage time of the process, i.e. $t_f$ itself
is random. More precisely, we consider a functional of the type
\begin{equation}
T= \int_0^{t_f} U\left(x(\tau)\right)\, d\tau
\label{fpf1}
\end{equation}
where $x(\tau)$ is a Brownian path starting from $x_0\ge 0$ at $\tau=0$ and propagating
up to time $\tau=t$ and $U(x)$, as before, is some specified function.
The integral in Eq.~(\ref{fpf1}) is up to the first-passage time $t_f$ which itself
is random in the sense that it varies from realization to realization of the Brownian path (see Fig.~(\ref{fig:fpf})). 
Such functionals 
appear in many problems (some examples are given
below) in physics, astronomy, queuing theory etc. and we will generally refer to 
them as first-passage Brownian functionals. 
\begin{figure}
\includegraphics[height=10.0cm,width=10.0cm,angle=0]{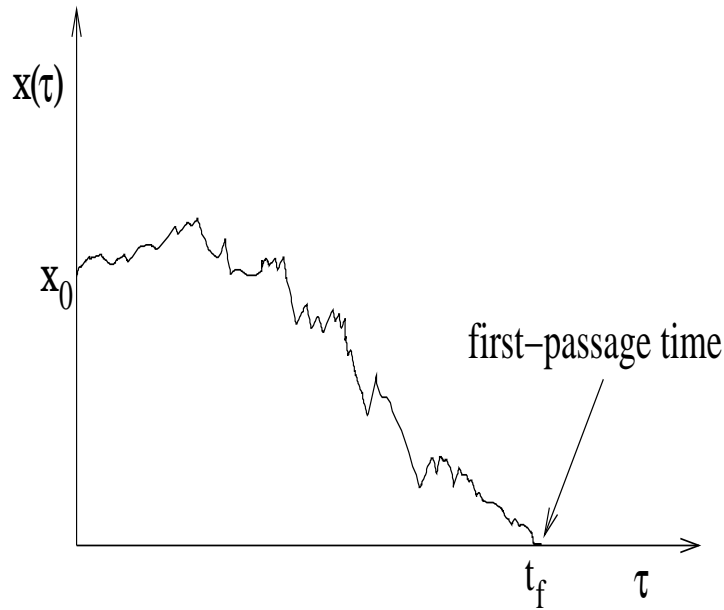}
\caption{\label{fig:fpf}
A Brownian path $x(\tau)$, starting at $x_0$ at $\tau=0$, crosses the origin for the first time
at $\tau=t_f$, $t_f$ being the first-passage time.}  
\end{figure}

We would like to compute the pdf $P(T|x_0)$ of $T$ in Eq.~(\ref{fpf1}) given that the
Brownian path starts at the initial position $x_0$. As before, it is useful to
consider the Laplace transform
\begin{equation}
Q(x_0)= \int_0^{\infty} e^{-p\,T} P(T|x_0) dT = \langle e^{-p \int_0^{t_f} U(x(\tau)) d\tau}\rangle
\label{fpf2}
\end{equation}
where the rhs is an average over all possible Brownian paths starting at $x_0$ at $\tau=0$
and stopping at the first time they cross the origin. For brevity, we have suppressed the $p$ 
dependence of $Q(x_0)$. 
Note that each path, starting from $x_0$, evolves via Eq.~(\ref{lange1}) where $\xi(t)$ is
a delta correlated white noise. Note also that $t_f$ varies from path to path.
Thus at first sight, this seems to be a rather difficult problem to solve. However, as we will
see now that in fact this problem is simpler than the previous problem over a fixed time interval
$[0,t]$! 

To proceed, we split a typical path over the interval $[0,t_f]$ into two parts: 
a left interval $[0,\Delta \tau]$ where the process proceeds from $x_0$ to $x_0+\Delta
x=x_0+\xi(0)\Delta \tau$ in a small time $\Delta \tau$ and a right interval $[\Delta \tau, t_f]$ in 
which
the process starts at $x_0+\Delta x$ at time $\Delta \tau$ and reaches $0$ at time $t_f$.
The integral $\int_0^{t_f}
U\left(x(\tau)\right)\,d\tau$ is also split into two parts: $\int_0^{t_f}=\int_0^{\Delta \tau} 
+\int_{\Delta 
\tau}^{t_f}$.
Since
the initial value is $x_0$, one gets $\int_0^{\Delta \tau}U\left(x(\tau)\right)\,d\tau= U(x_0)
\Delta \tau$ for small $\Delta \tau$. Then the Eq.~(\ref{fpf2}) can be written as
\begin{equation}
Q(x_0)=\langle e^{-p\int_0^{t_f} U\left(x(\tau)\right)\,d\tau}\rangle = 
{\langle e^{-p\,U(x_0)\Delta \tau} Q(x_0+\Delta x)\rangle}_{\Delta x},
\label{T2}
\end{equation}
where we have used the fact that for the right interval $[\Delta \tau, t_f]$, the starting position
is $x_0+\Delta x=x_0+\xi(0)\Delta \tau$, which itself is random. The average in the second line
of Eq.~(\ref{T2}) is over all possible realizations of $\Delta x$. We then substitute $\Delta
x=\xi(0)\Delta \tau$ in Eq.~(\ref{T2}), expand in powers of $\Delta \tau$ and
average over the noise $\xi(0)$. We use the fact that the noise $\xi(t)$ is delta correlated, i.e.
$\langle \xi^2(0)\rangle = 1/{\Delta \tau}$ as $\Delta \tau\to 0$. The leading order term on the right
hand side of Eq.~(\ref{T2}) is independent of $\Delta \tau$ and is simply $Q(x_0)$ which
cancels the same term on the left hand side of Eq.~(\ref{T2}). Collecting the rest of
the terms we get 
\begin{equation}
\left[ \frac{1}{2} \frac{d^2 Q}{dx_0^2} - p\,U(x_0)Q(x_0)\right]\Delta \tau
+O\left( (\Delta \tau)^2\right) =0.
\label{T3}
\end{equation}
Equating the leading order term to zero provides us an ordinary differential equation
\begin{equation}
\frac{1}{2} \frac{d^2 Q}{dx_0^2} - p\,U(x_0)Q(x_0)=0
\label{T4}
\end{equation}
which is valid in $x_0\in [0,\infty]$ with the following boundary conditions:
(i) When the initial position $x_0\to 0$, the first-passage time $t_f$ must also be $0$. Hence
the integral $\int_0^{t_f}U\left(x(\tau)\right)\,d\tau=0$. From the definition in Eq.~(\ref{T2}), 
we get
$Q(x_0=0)=1$ and (ii) when the initial position $x_0\to \infty$, the first-passage time
$t_f\to \infty$, hence the integral $\int_0^{t_f}U\left(x(\tau)\right)\,d\tau$ also diverges in 
this limit, at least
when $U(x)$ is a nondecreasing function of $x$.
The definition in Eq.~(\ref{T2}) then gives the boundary condition, $Q (x_0\to \infty)=0$.

So, given a functional $U(x)$, the scheme would be to first solve the the ordinary differential equation 
(\ref{T4}) with the appropriate boundary conditions mentioned above to obtain $Q(x_0)$
explicitly and then invert the Laplace transform in Eq.~(\ref{fpf2}) to get the desired
pdf $P(T|x_0)$ of the first-passage functional. As a simple test of this method, let us
first consider the case $U(x)=1$. In this case the functional 
$T=\int_0^{t_f} U\left(x(\tau)\right)\,d\tau=t_f$
is the first-passage time itself. The differential equation (\ref{T4}) can be trivially solved 
and the solution satisfying the given boundary conditions is simply
\begin{equation}
Q(x_0)= e^{-\sqrt{2p}\, x_0}.
\label{T5}
\end{equation}
Inverting the Laplace transform with respect to $p$ gives the pdf of the first-passage time
\begin{equation}
P(t_f|x_0)= \frac{x_0}{\sqrt{2\pi}}\, \frac{e^{-x_0^2/{2t_f}}}{{t_f}^{3/2}},
\label{T6}
\end{equation}
which is identical to the result in Eq.~(\ref{fp4}) obtained by the path integral method.
Below we provide a few nontrivial examples and applications of this method.

\subsection{ Area till the first-passage time}

Here we calculate the pdf of the area under a Brownian motion (starting at $x_0$) till
its first-passage time~\cite{KM}. Thus the relevant functional is $A=\int_0^{t_f} 
x(\tau)d\tau$
and hence $U(x)=x$. In Fig.~(\ref{fig:fpf}), $A$ is just the area under the curve over the
time interval $[0,t_f]$. This problem has many 
applications in combinatorics and queuing theory.
For example, an important object in combinatorics is the area of a lattice polygon in
two dimensions~\cite{Kear1}.
A particular example of a lattice polygon is the rooted staircase polygon whose two
arms can be thought of as two independent random walkers whose trajectories meet for the first time at 
the end of the polygon. The difference walk between these two arms then defines, in the continuum limit,
a Brownian motion.
The area of such a polygon can then be approximated, in the continuum
limit, by the area under a single Brownian motion till its first-passage time~\cite{KM}.  
This picture also relates this problem to the directed Abelian sandpile model~\cite{DRam}
where $t_f$ is just the avalanche duration and the area $A$ is the size of an 
avalanche cluster. Another application arises in queueing theory, where the length of a queue 
$l_n$ after $n$ time steps evolves stochastically~\cite{Kear1}. In the simplest approximation, one considers 
a random walk model, $l_n = l_{n-1} +\xi_n$ where $\xi_n$'s are independent and identically
distributed random variables which model the arrival and departure of new customers.
When the two rates equal, $\langle \xi_n\rangle=0$. 
In the large $n$ limit, $l_n$ can be approximated by a Brownian motion $x(\tau)$, whereupon
$t_f$ becomes the so called `busy' period (i.e. the time until the queue first becomes empty)
and the area $A$ then approximates the total number of customers served during the busy period. 

Substituting $U(x)=x$ in Eq.~(\ref{T4}), one can solve the differential equation with
the prescribed boundary conditions and the solution is~\cite{KM}
\begin{equation}
Q(x_0)= 3^{2/3} \Gamma(2/3) {\rm Ai} (2^{1/3} p^{1/3} x_0)
\label{A1}
\end{equation}
where ${\rm Ai}(z)$ is the Airy function. It turns out that this Laplace transform can be 
inverted to give an explicit expression for the pdf~\cite{KM}
\begin{equation}
P(A|x_0)= \frac{2^{1/3}}{3^{2/3} \Gamma(1/3)}\, \frac{x_0}{A^{4/3}}\, \exp\left[-\frac{2x_0^3}{9A}\right].
\label{A2}
\end{equation}
Thus the pdf has a power law tail for large $A\gg x_0^3$, $P(A|x_0)\sim A^{-4/3}$ and
an essential singularity $P(A|x_0)\sim \exp[-2x_0^3/{9A}]$ for small $A\to 0$. 
Following the same techniques, one can also derive the pdf of the area till the first-passage time under 
a Brownian motion    
with a drift towards the origin--in this case the pdf has a stretched exponential tail
for large $A$~\cite{KM}, $P(A|x_0)\sim A^{-3/4}\exp[-\sqrt{8\mu^3 A/3}]$ where $\mu$ is the drift.

Note the difference between the pdf of the area $P(A|x_0)$, under a Brownian motion till its first-passage time
starting at $x_0$ at $\tau=0$, as given in Eq.~(\ref{A2}) and the pdf of the area under a Brownian
excursion $P(A,t)$ in Eq.~(\ref{area2}). In the latter case, the Brownian path is conditioned to
start at $x_0=0$ at $\tau=0$ and end at $x=0$ at $\tau=t$ and one is interested in the statistics
of the area under such a conditioned path over the {\em fixed} time interval $t$. In the former
case on the other hand, one is interested in the area under a free Brownian motion starting
at $x_0>0$ and propagating up to its first-passage time $t_f$ that is {\em not fixed} but
varies from one realization of the path to another.

\subsection{Time period of oscillation of an undamped particle in a random potential}

The study of transport properties in a system with quenched disorder is an important area
of statistical physics~\cite{BG}. The presence of a quenched disorder makes analytical calculations
hard and very few exact results are known. Perhaps the simplest model that captures some
complexities associated with the transport properties in disordered systems is that
of a classical Newtonian particle moving in a one dimensional random potential $\phi(x)$
\begin{equation}
m\frac{d^2x}{dt^2} + \Gamma \frac{dx}{dt} = F\left(x(t)\right) + \xi(t)
\label{sinai0}
\end{equation}
where $F(x)= -d\phi/dx$ is the force derived from the random potential $\phi(x)$, $\Gamma$
is the friction coefficient and $\xi(t)$ is the thermal noise with zero mean
and a delta correlator, $\langle \xi(t)\xi(t')\rangle= 2D \delta(t-t')$ with
$D=k_B T/\Gamma$ by the Stokes-Einstein relation (\ref{e3}). 

It turns out that even this 
simple problem is very hard to solve analytically for an arbitrary random potential
$\phi(x)$. A special choice of the random potential where one can make some progress  
is the Sinai potential~\cite{Sinai}, where one assumes that 
$\phi(x)=\int_0^{x} \eta(x') dx'$. The variables $\eta(x)$'s have zero mean
and are delta correlated $\langle \eta(x_1)\eta(x_2)\rangle= \delta(x_1-x_2)$. Thus the
potential $\phi(x)$ itself can be considered as a Brownian motion in space. 
In the overdamped limit when the frictional force is much larger than the
inertial force, Eq.~(\ref{sinai0}) then reduces to the Sinai model~\cite{Sinai}
\begin{equation}
\Gamma \frac{dx}{dt} = F(x=x(t)) +\xi(t)
\label{sinai1}
\end{equation}
where the random force $F(x)=-d\phi/dx = \eta(x)$ is just a delta correlated white noise
with zero mean: $\langle F(x)\rangle=0$ and $\langle F(x)F(x')\rangle= \delta(x-x')$.

\begin{figure}
\includegraphics[height=10.0cm,width=10.0cm,angle=0]{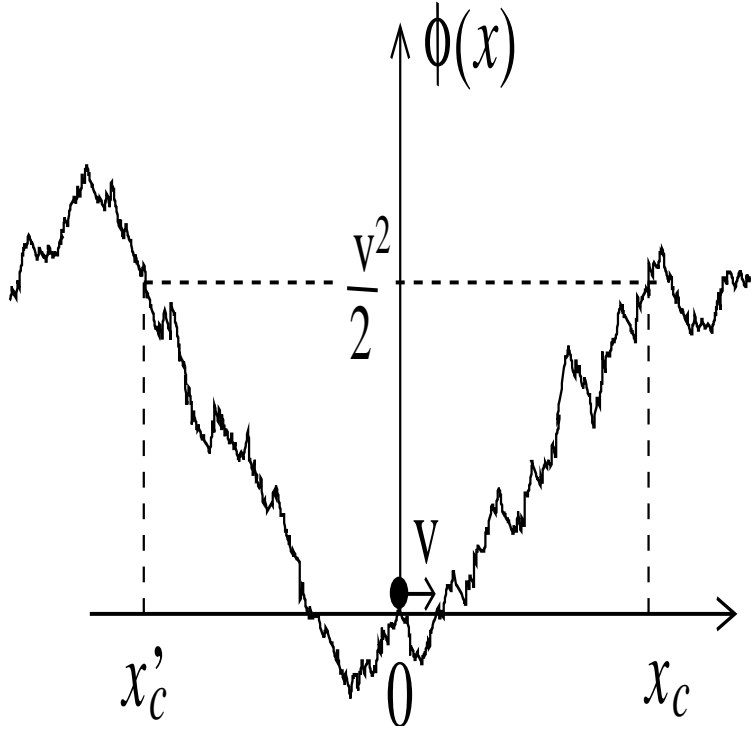}
\caption{\label{fig:newton}
A Newtonian particle in a Brownian potential with initial velocity $v$. The right/left 
turning points are shown as $x_c$ and $x_c'$ respectively where the potential energy first
becomes equal to the kinetic energy, $\phi=v^2/2$.} 
\end{figure}

Here we consider a simple model~\cite{DM1} where the particle diffuses in the
same Sinai potential $\phi(x)=\int_0^{x} \eta(x') dx'$, but we consider 
the opposite limit where the particle is {\em undamped}, i.e. $\Gamma=0$
and is driven solely by the inertial force. For simplicity, we also
consider the zero temperature limit where the thermal noise term drops out
of Eq.~(\ref{sinai0}) as well and one simply has
\begin{equation}
m \frac{d^2x}{dt^2} = F(x=x(t))
\label{new1}
\end{equation}
where $F(x)$ is a same random Sinai force as mentioned above.
We set $m=1$ and assume that the particle starts at the origin
$x=0$ with initial velocity $v>0$. Thus the particle will move to the right till it reaches 
a turning point $x_c$ where the potential energy becomes equal to the kinetic energy, i.e.
$\phi(x_c)= v^2/2$ and then it will move back to $x=0$ with a velocity $-v$ (see 
Fig.~(\ref{fig:newton})).
After returning to the origin with velocity $-v$, the particle will go to the left till it encounters
the first turning point on the left of the origin where it will turn and then will return to
the origin. Let $T$ and $T'$ denote the time for the particle to go from the origin to the
turning point at the right and to the one at the left respectively. Thus the particle will 
oscillate between the two turning points nearest to the origin
on either side and the time period of oscillation is $T_{\rm osc}= 2(T+T')$.       
Note that the variables $T$ and $T'$ will vary from one sample of quenched disorder
to another. The goal is to compute the probability distribution of $T$ and $T'$
and hence that of $T_{\rm osc}$. Since $\phi(x)$ is a Brownian motion in $x$, it follows from its 
Markov property that $\phi(x)$ for $x>0$ and for $x<0$ are completely independent of each other.
Thus $T$ and $T'$ are also independent and by symmetry, have identical distributions.
The distribution of $T_{\rm osc}$ can then be easily calculated by convolution.

To compute the pdf $P(T)$ of $T$ (starting at $x_0=0$), we first express $T$ as a functional of the 
Brownian potential
\begin{equation}
T = \int_0^{x_c} \frac{dx}{\sqrt{v^2-2\phi(x)}}
\label{new2}
\end{equation}
where $x_c$ is defined as the point where $\phi(x_c)=v^2$. On identifying the space
as the time $x\equiv \tau$ and the random potential $\phi$ as the trajectory of a random walk in 
space $x$, i.e.  $\phi \leftrightarrow x, \,\, x\leftrightarrow \tau$, $T$ in
Eq.~(\ref{new2})
is of the general form in Eq.~(\ref{fpf1}) with $U(x)= 1/\sqrt{v^2-2x}$  
and $x_c=t_f$ denoting the first-passage time to the level $x=v^2/2$,
starting at $x_0$. Following the general scheme, we need to solve the
differential equation (\ref{T4}), now valid for $ -\infty \le x_0 \le v^2/2$,
with $U(x)=1/\sqrt{v^2-2x}$ and the boundary conditions, $Q(x_0\to -\infty)=0$
and $Q(x_0\to v^2/2)=1$. Upon finding the solution one needs to put $x_0=0$
and then invert the Laplace transform. This can be done explicitly and one 
obtains~\cite{DM1} 
\begin{equation}
P(T) = \frac{2^{2/3} v^2}{3^{4/3} \Gamma(2/3)}\, \frac{1}{T^{5/3}}\, \exp\left[-\frac{2v^3}{9T}\right].
\label{new3}
\end{equation}
This is one of the rare examples of an exact result on a transport property
in a quenched disorderd system, thus illustrating the power of 
the approach outlined in this section.

\subsection{Distribution of the Lifetime of a Comet in Solar System}

In this final subsection we provide an example from astrophysics~\cite{Ham} where the general
technique of the first-passage Brownian functional is applicable. A comet enters a
solar system with a negative energy $E_0<0$ and keeps orbiting around the sun
in an elliptical orbit whose semimajor axis length $a$ is determined by the
relation $E_0= - GM/{2a}$ where $G$ is the gravitational constant and $M$ is
the mass of the sun. It was first pointed out by Lyttleton~\cite{Lyttle} that
the energy of the comet gets perturbed by Jupiter each time the comet visits 
the neighbourhood of the sun and the planets and successive perturbations lead
to a positive energy of the comet which then leaves the solar system. It is
convenient to work with the negative energy $x=-E>0$ of the comet. We assume that
the comet enters the solar system 
with initial negative energy $x_0$ and 
has values of $x$ equal to $x_1$, $x_2$, $\ldots$, $x_{t_f}$ at successive
orbits till the last one labelled by $t_f$ when its value of $x$ crosses $0$
(energy becomes positive) and it leaves the solar system. The
lifetime of the comet is given by
\begin{equation}
T= U(x_0) + U(x_1) +\ldots U(x_{t_f})
\label{ham1}
\end{equation}
where $U(x)$ is the time taken to complete an orbit with negative energy $x>0$.
According to Kepler's third law, $U(x) = c\, x^{-3/2}$ where $c$ is an constant
which we set to $c=1$ for convenience.
Moreover, a simple way to describe the perturbation due to Jupiter
is by a random walk model, $x_n = x_{n-1} + \xi_n$ where $\xi_n$ is the noise
induced by Jupiter and is assumed to be independent from orbit to orbit~\cite{Ham}. 
Within this random walk theory, the lifetime of a comet in Eq.~(\ref{ham1}), 
in the continuum limit becomes a first-passage Brownian functional~\cite{Ham}
\begin{equation}
T= \int_0^{t_f} [x(\tau)]^{-3/2} d\tau
\label{ham2}
\end{equation}
where the random walk starts at $x_0$ and ends at its first-passage time $t_f$ when
it first crosses the origin. The pdf $P(T|x_0)$ was first obtained by Hammersley~\cite{Ham}.
Here we show how to obtain this result using the general approach outlined here
for first-passage Brownian functionals.

Following our general scheme, we thus 
have $U(x)= x^{-3/2}$ in the differential Eq.~(\ref{T4}). The solution, satisfying the
proper boundary conditions, can be easily found
\begin{equation}
Q(x_0)= 16 p {x_0}^{1/2} K_2\left( \sqrt{32 p}\, {x_0}^{1/4}\right)
\label{ham3}
\end{equation} 
where $K_2(z)$ is the modified Bessel function of degree $2$. Next, we need to invert the
Laplace transform in Eq.~(\ref{ham3}) with respect to $p$. This can be done by using
the following identity
\begin{equation}
\int_0^{\infty} y^{-\nu-1} e^{-p y -\beta/y} dy = 2 {\left(\frac{p}{\beta}\right)}^{\nu/2} 
K_{\nu}(2\sqrt{\beta p}).
\label{ham4}
\end{equation}
Choosing $\beta= 8\sqrt{x_0}$, we can invert the laplace transform to obtain the
exact pdf $P(T|x_0)$ of the lifetime of a comet
\begin{equation}
P(T|x_0)= \frac{64 x_0}{T^{3}}\, \exp\left[-\frac{8\sqrt{x_0}}{T}\right].
\label{ham5}
\end{equation}

It is worth pointing out that in all three examples above, the pdf $P(T|x_0)$ of the
first-passage Brownian functional has a power law tail $P(T|x_0)\sim T^{-\gamma}$ for
large $T$ and and an essential singularity in the limit $T\to 0$. While the exponent
of the power law tail can be easily obtained using a scaling argument, the essential
singular behavior at small $T$ is not easy to obtain just by a scaling argument. 

\section{Conclusion}

In this article I have provided a brief and pedagogical review of the techniques
to calculate the statistical properties of functionals of one dimensional Brownian motion.
It also contains a section devoted to `first-passage' Brownian functional, a quantity
that appears in many problems but the techniques to calculate its properties are somewhat
less known compared to the standard Feynman-Kac formalism for the usual Brownian functional.
A simple backward Fokker-Planck approach is provided here to calculate the probability
distribution of a first-passage Brownian functional.
Several examples and applications of the standard Brownian functionals as well as the
first-passage Brownian functionals from physics, 
probability theory, astronomy and in particular
from computer science are provided. 

The techniques detailed in this article are valid for free Brownian motion in one dimension.
However, they can be easily generalized to study the functionals of a Brownian motion
in an external potential. The external potential can represent e.g. a constant
drift~\cite{CMY,CT,KM} or a harmonic potential~\cite{MB}. Alternately, the external
potential can be random as in a disordered system. The backward Fokker Planck approach
reviewed here has been particularly useful in calculating exactly the disorder averaged
distributions of Brownian functionals in the Sinai model~\cite{CMY,occup1,pers1}.

There are several open directions for future research. For example, to the best of my knowledge,
the properties of first-passage Brownian functionals have so far not been studied in 
disordered systems. The techniques discussed here could be useful in that direction.
Though there have been few studies of Brownian functionals in higher dimensions, there
are still many open problems with direct relation to experiments~\cite{CDT} and
more studies in that direction would be welcome. Finally, the discussion in this
article is limited to the simple
Brownian motion which is a Gaussian as well as a Markov process. In many
real systems, the relevant stochastic process often is non-Gaussian and/or non-Markovian.
It would certainly be interesting to study the properties of functionals of such
stochastic processes. 

In summary, I hope I have been able to convey to the reader the beauty and the interests 
underlying Brownian `functionalogy' with its diverse applications ranging from
physics and astronomy to computer science, making it a true legacy of Albert
Einstein whose 1905 paper laid the basic foundation of this interesting subject.

\vspace{0.4cm}

\noindent{\bf Acknowledgements:} It is a pleasure to thank my collaborators A.J. Bray, A. Comtet, C. 
Dasgupta, D.S. Dean, A. Dhar, M.J. 
Kearney, and S. Sabhapandit with whom I have worked on Brownian functionals in the recent past.
I also acknowledge useful discussions on Brownian motion and related topics 
with M. Barma, J. Desbois, D. Dhar, P.L. Krapivsky, S. Redner, C. Sire, C. Texier, M. Yor and R.M. Ziff.

\end{document}